\documentclass[twocolumn,pre,superscriptaddress]{revtex4}

\usepackage{dcolumn}
\usepackage{amsmath}

\usepackage{graphicx}

\newcommand{\dd}{\mathrm{d}}
\newcommand{\ii}{\mathrm{i}}

\newcommand{\av}[1]{\left\langle#1\right\rangle}
\newcommand{\etal}{{\it{}et~al.}}
\newcommand{\defn}{\textit}
\newcommand{\Ord}{\mathrm{O}}
\newcommand{\Tr}{\mathop\mathrm{Tr}}
\renewcommand{\Im}{\mathop\mathrm{Im}}
\newcommand{\mat}{\mathbf}

\begin{document}

\title{Spectra of random networks with arbitrary degrees}
\author{M. E. J. Newman}
\affiliation{Department of Physics, University of Michigan, Ann Arbor, Michigan, USA}
\affiliation{Center for the Study of Complex Systems, University of Michigan, Ann Arbor, Michigan, USA}
\author{Xiao Zhang}
\affiliation{Department of Physics, University of Michigan, Ann Arbor, Michigan, USA}
\author{Raj Rao Nadakuditi}
\affiliation{Department of Electrical Engineering and Computer Science,
  University of Michigan, Ann Arbor, MI 48109}

\begin{abstract}
  We derive a message passing method for computing the spectra of locally tree-like networks and an approximation to it that allows us to compute closed-form expressions or fast numerical approximates for the spectral density of random graphs with arbitrary node degrees---the so-called configuration model.  We find the latter approximation to work well for all but the sparsest of networks.  We also derive bounds on the position of the band edges of the spectrum, which are important for identifying structural phase transitions in networks.
\end{abstract}

\maketitle

\section{Introduction}
The spectral properties of the adjacency matrix of a network play a central role in the analysis of network structure, for instance in the eigenvector centrality~\cite{Bonacich87}, in graph partitioning and community detection~\cite{Fiedler73,PSL90,Newman06b}, in the theory of dynamical systems on networks~\cite{PG16}, and in the analysis of structural phase transitions such as percolation~\cite{BBCR10,KNZ14}, localization~\cite{CM11,MZN14}, and detectability transitions~\cite{DKMZ11a,NN12}.  

There exist well-known computer algorithms for calculating the eigenvalue spectrum of matrices, which can be applied directly to the adjacency matrix of a network, but they are numerically demanding, taking time of order~$n^3$ to calculate all eigenvalues of an $n$-node network.  Furthermore, they give only the spectrum of a single network instance, where in many cases we would like to calculate the average spectrum of an entire graph ensemble.  In this paper we develop an alternative approach to calculating network spectra, based on message passing methods and focusing particularly on the case of the so-called configuration model---a random graph with given degree sequence and one of the standard models in the theory of complex networks.  We first develop a general message passing method, which is exact on arbitrary networks that are free of short loops and which is closely related to previous approaches for calculating graph spectra.  Then we further develop an approximation to this method for the particular case of the configuration model in the limit of large network size that allows us to find closed-form expressions for the spectral density in certain cases and to perform numerical calculations in $\Ord(1)$ time in others.

There has been a significant amount of previous work on the spectra of complex networks.  In early work, Farkas~\etal~\cite{FDBV01} computed spectra for a range of networks using standard numerical methods and demonstrated clear deviations of the spectral density from the Wigner form expected for traditional dense random graphs.  Goh~\etal~\cite{GKK01a} looked at networks with power-law degree distributions (a common feature of many empirical networks), giving both numerical results and analytic bounds for the largest eigenvalues.  Dorogovtsev~\etal~\cite{DGMS03} gave an analytic prescription for calculating complete spectra of configuration model networks but were unable to solve the resulting equations.  Instead therefore, they develop an approximation, which is similar in some respects to our own though different in both motivation and final form, and which appears to give good results in some cases but is rather inaccurate in others, such as networks with power-law degree distributions.  A~contrasting approach has been pursued by Semerjian and Cugliandolo~\cite{SC02b} who start from the classic formulation of Edwards and Jones~\cite{EJ76} of the spectral density in terms of a Gaussian path integral and use a replica-type analysis to derive several different approximations to the spectra of a traditional Erd\H{o}s--R\'enyi style random graph.  K\"uhn~\cite{Kuhn08} applied similar methods to the configuration model, using an analysis reminiscent of the Viana--Bray solution for a dilute spin-glass to derive an approximation to the spectrum that is similar in its accuracy to ours, although quite technically daunting.  Rogers~\etal~\cite{RCKT08}, starting again from the Edwards--Jones formulation, derive a message passing method essentially equivalent to the one we use, which they employ as a numerical tool for computing spectra.  Chung~\etal~\cite{CLV03} study a slightly different class of model networks, those with given expected degrees and statistically independent edges (sometimes called the Chung--Lu model following earlier work by some of the same authors~\cite{CL02b}), for which they derive an expression for the single largest eigenvalue, which plays a role for instance in percolation calculations~\cite{BBCR10}.  Using tools from free probability theory, Nadakuditi and Newman~\cite{NN13} calculated complete spectra for the same class of networks.

In outline this paper is as follows.  In Section~\ref{sec:spectrum} we derive message passing equations for the spectral density which form the basis for subsequent developments.  In Section~\ref{sec:approximation} we derive our approximation to the message passing equations for the case of the configuration model and show how it can be used both to perform fast numerical calculations and, in some cases, to give analytic solutions for the spectral density.  We also use it to derive bounds on the position of the edges of the spectrum, which play a central role in the theory of structural phase transitions in networks.  In Section~\ref{sec:examples} we give a number of example applications of our methods, showing the accuracy of the approach in some cases, as well as other cases where it breaks down.  In Section~\ref{sec:concs} we give our conclusions and suggest some directions for future work.

\section{Spectral density of a locally tree-like network}
\label{sec:spectrum}
Our calculations begin with the derivation of a system of message passing equations for computing the spectral density of a locally tree-like network, meaning any network where the local neighborhood of any node, out to any fixed distance, takes the form of a tree with probability one in the limit of large network size.  This part of the calculation is similar to developments described previously by Dorogovtsev~\etal~\cite{DGMS03} and Rogers~\etal~\cite{RCKT08}, although our derivation is interesting in its own right because it employs only elementary algebraic methods, where previous approaches have relied on heavier mathematical machinery.

Suppose we are given a single undirected unweighted network of $n$ nodes and asked to calculate its spectral density, which is the function
\begin{equation}
\rho(x) = {1\over n} \sum_{i=1}^n \delta(x - \lambda_i),
\label{eq:rho1}
\end{equation}
where $\delta(x)$ is the Dirac delta function and $\lambda_1\ldots\lambda_n$ are the $n$ eigenvalues of the adjacency matrix~$\mat{A}$ of the network---the matrix with binary-valued elements~$A_{uv}$ equal to 1 whenever there is an edge between nodes~$u,v$ and 0 otherwise.  Following a standard line of development we express the delta function as the limit of a Lorenzian (or Cauchy) distribution:
\begin{equation}
\delta(x) = \lim_{\epsilon\to0^+} {\epsilon/\pi\over x^2+\epsilon^2}
  = -{1\over\pi} \lim_{\epsilon\to0^+} \Im {1\over x+\ii\epsilon},
\label{eq:delta}
\end{equation}
where the notation $\lim_{\epsilon\to0^+}$ means that the parameter~$\epsilon$, which controls the width of the Lorenzian, tends to zero from above.  Substituting \eqref{eq:delta} into~\eqref{eq:rho1}, we find that
\begin{equation}
\rho(x) = -{1\over n\pi} \lim_{\epsilon\to0^+} \Im \sum_{i=1}^n
           {1\over x-\lambda_i+\ii\epsilon}.
\end{equation}
It will be convenient for subsequent developments to generalize this spectral density into the complex plane, defining $z = x+\ii\epsilon$ and
\begin{equation}
\rho(z) = -{1\over n\pi} \sum_{i=1}^n {1\over z-\lambda_i}
        = -{1\over n\pi} \Tr (z\mat{I}-\mat{A})^{-1},
\label{eq:complexrho}
\end{equation}
where $\mat{I}$ is the identity.  The standard spectral density for the network is the limiting value of the imaginary part of this function as $z$ tends to the real line from above.  In fact, in many practical situations it is desirable to retain a small imaginary part for~$z$, corresponding to $\epsilon>0$, producing a Lorenzian broadening of the delta-function peaks in the spectral density.  For finite networks this gives us a smooth density function~$\rho(z)$ rather than a set of spikes, effectively making a kernel density estimate of the spectral density with a Lorenzian kernel.

If we expand the matrix inverse in~\eqref{eq:complexrho} as a geometric series $(z\mat{I}-\mat{A})^{-1} = z^{-1} \sum_{k=0}^\infty (\mat{A}/z)^k$ and take the trace term by term, we find that
\begin{equation}
\rho(z) = -{1\over n\pi z} \sum_{k=0}^\infty {\Tr \mat{A}^k\over z^k}.
\label{eq:rho2}
\end{equation}
The quantity~$\Tr \mat{A}^k$ is equal to the number of closed walks of length~$k$ in the network---paths that start at any node and return there (not necessarily for the first time) exactly $k$ steps later.  If we can count such closed walks on our network for all values of $k$ then we can compute the spectral density from Eq.~\eqref{eq:rho2}.  We do this counting using a message passing method.

\subsection{Message passing}
\label{sec:messages}
As we have said, our focus here is on locally tree-like networks, meaning networks in which local neighborhoods are trees, having a vanishing density of short loops.  The absence of loops means that any closed walk must necessarily start and end by traversing the same edge---it cannot return to the starting node by any edge other than the one it left by, since in so doing it would complete a loop in the network, of which there are none.  Indeed \emph{every} edge in a closed walk on a locally tree-like network must, for the same reason, be traversed twice, once in each direction, or more generally the same number of times in both directions.  This in turn means that all closed walks have an even number of steps.

Let $n^{uv}_{2r}$, with $r$ a positive integer, be the number of closed walks that begin by traversing the edge from $u$ to~$v$ and end, after exactly $2r$ steps, by traversing the same edge back again from $v$ to~$u$ for the first time.  Other edges may be traversed any (even) number of times, but the edge between $u$ and $v$ is traversed only once each way.

The smallest possible value of $r$ in this scenario is~1, for which $n^{uv}_2 = 1$ trivially.  For all higher values $r>1$, we can write a self-consistent expression for $n^{uv}_{2r}$ thus:
\begin{align}
n^{uv}_{2r} &= \sum_{m=1}^\infty \biggl[ \sum_{\substack{w_1\in N_v\\w_1\ne u}}
  \!\ldots\!\sum_{\substack{w_m\in N_v\\w_m\ne u}} \biggr]
  \biggl[ \sum_{r_1=1}^\infty \ldots \sum_{r_m=1}^\infty \biggr]
  \prod_{i=1}^m n^{vw_i}_{2r_i} \nonumber\\
  &\hspace{4em}{}\times \delta\biggl( r-1,\sum_{i=1}^m r_i \biggr),
\label{eq:nij}
\end{align}
where $N_v$ denotes the set of neighbor nodes of~$v$ and $\delta(i,j)$ is the Kronecker delta.  To break it down, this expression says that a walk starting and ending along the edge $uv$ makes some number~$m$ of subsequent excursions from node~$v$ each of which has a first step to one of the neighbors of~$v$ other than~$u$, that the total number of such walks is the product of the numbers of distinct excursions, and that the individual lengths $2r_1\ldots2r_m$ of these excursions necessarily sum to $2r-2$.

To solve this system of equations for the numbers~$n^{uv}_{2r}$, we define the useful quantity
\begin{equation}
h^{uv}(z) = \sum_{r=1}^\infty {n^{uv}_{2r}\over z^{2r}}.
\label{eq:defshuv}
\end{equation}
Substituting for $n^{uv}_{2r}$ from Eq.~\eqref{eq:nij} into Eq.~\eqref{eq:defshuv} and performing the sum over~$r$, we get
\begin{align}
h^{uv}(z)
  &= {1\over z^2} \sum_{m=1}^\infty 
     \biggl[ \sum_{\substack{w_1\in N_v\\w_1\ne u}}\ldots
     \sum_{\substack{w_m\in N_v\\w_m\ne u}} \biggr]
     \prod_{i=1}^m \,\sum_{r_i=1}^\infty {n^{vw_i}_{2r_i}\over z^{2r_i}}
     \nonumber\\
  &= {1\over z^2} \sum_{m=1}^\infty \,\prod_{i=1}^m\,
     \sum_{\substack{w_i\in N_v\\w_i\ne u}} h^{vw_i}(z) \nonumber\\
  &= {1\over z^2} \sum_{m=1}^\infty \biggl[ \sum_{\substack{w\in N_v\\w\ne u}}
     h^{vw}(z) \biggr]^m.
\end{align}
The remaining sum over~$m$ is a simple geometric series, which can be completed to give
\begin{equation}
h^{uv}(z) = {1/z^2\over 1 - \sum_{\substack{w\in N_v\\w\ne u}} h^{vw}(z)}.
\label{eq:messages}
\end{equation}
This is our fundamental message passing equation.  We can think of $h^{uv}(z)$ as a message passed from node~$v$ to its neighbor~$u$, whose value can be computed from the values of the messages received by~$v$ from its other neighbors~$w$.

If one can compute the values of the messages~$h^{uv}(z)$ for any given value of~$z$, one can compute the spectral density itself as follows.  By analogy with Eq.~\eqref{eq:nij}, the number~$n^u_{2r}$ of closed walks of length~$2r$ that start and end at node~$u$ can be written as
\begin{align}
n^u_{2r} &= \sum_{m=1}^\infty \biggl[ \sum_{v_1\in N_u}\ldots \sum_{v_m\in N_u}
            \biggr]
  \biggl[ \sum_{r_1=1}^\infty \ldots \sum_{r_m=1}^\infty \biggr]
  \prod_{i=1}^m n^{uv_i}_{2r_i} \nonumber\\
  &\hspace{4em}{}\times \delta\biggl( r,\sum_{i=1}^m r_i \biggr).
\label{eq:ni}
\end{align}
Then we define
\begin{align}
g^u(z) &= \sum_{r=1}^\infty {n^u_{2r}\over z^{2r}}
        = \sum_{m=1}^\infty \biggl[ \sum_{v_1\in N_u}\ldots \sum_{v_m\in N_u}
          \biggr] \prod_{i=1}^m \sum_{r_i=1}^\infty {n^{uv_i}_{2r_i}\over
          z^{2r_i}} \nonumber\\
       &= \sum_{m=1}^\infty \prod_{i=1}^m \sum_{v_i\in N_u} h^{uv_i}(z)
        = \sum_{m=1}^\infty \biggl[ \sum_{v\in N_u} h^{uv}(z)  \biggr]^m,
\end{align}
and hence
\begin{equation}
g^u(z) = {1\over1 - \sum_{v\in N_u} h^{uv}(z)}.
\label{eq:gu}
\end{equation}
The spectral density, Eq.~\eqref{eq:rho2}, can now be written in terms of this quantity as
\begin{equation}
\rho(z) = -{1\over n\pi z} \sum_{u=1}^n g^u(z)
        = -{1\over n\pi z} \sum_{u=1}^n {1\over1 - \sum_{v\in N_u} h^{uv}(z)}.
\label{eq:rho3}
\end{equation}

Between them, Eqs.~\eqref{eq:messages} and~\eqref{eq:rho3} give us our prescription for calculating the spectral density.  Note that the variable~$z$ enters Eq.~\eqref{eq:messages} only as~$z^2$, which means that the spectral density will always be symmetric about the origin.

The message passing equations can be used as a numerical tool for computing network spectra, as Rogers~\etal~\cite{RCKT08} do with the equivalent equations they derive.  One simply chooses initial values of the messages (for instance at random) and iterates Eq.~\eqref{eq:messages} repeatedly until convergence is achieved.  Direct solution of the message passing equations, however, will not be our primary goal here.

\section{Degree-based approximation}
\label{sec:approximation}
The developments presented so far provide an elementary derivation of a message passing method for calculating network spectra.  Though interesting, however, the method derived performs essentially the same calculation as the previously proposed method of Rogers~\etal~\cite{RCKT08} and in this sense does not add much to our toolkit.  In this section, however, we go further, focusing specifically on the configuration model and introducing an approximation to the message passing algorithm that reduces its accuracy very little while making it enormously faster and, in some cases, allowing us to compute analytic solutions for the spectral density.

The configuration model~\cite{MR95,NSW01}, a random graph model with arbitrary node degrees, is one of the most fundamental of network models.  To generate a configuration model network one fixes the degree of each node separately, then connects nodes at random while respecting the degrees.  The configuration model is widely used both for understanding network structure in general and as a starting point for further calculations of network properties and processes.

\begin{figure}
\begin{center}
\includegraphics[width=6cm]{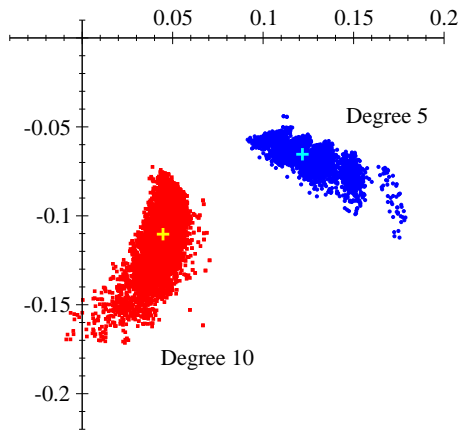}
\end{center}
\caption{The values of the messages $h^{uv}(z)$ at $z = 3 + \ii\epsilon$ with $\epsilon=0.01$, plotted in the complex plane, for a configuration model network with nodes of two different degrees: half the nodes have degree~5 and half have degree~10.  The two clouds of points correspond to nodes~$v$ of the two different degrees as indicated.  The plus signs mark the mean or centroid of each cloud.  Our approximation consists of replacing the value of each message with the centroid for the cloud it belongs to.}
\label{fig:messages}
\end{figure}

The configuration model generates networks that are locally tree-like and hence the message passing approach can be applied to them.  We do this in Fig.~\ref{fig:messages} for an example network in which nodes have just two distinct degrees, 5~and~10, with equal probability.  The figure shows a scatter plot in the complex plane of the resulting values of the messages~$h^{uv}(z)$ at the point $z = 3 + \ii\epsilon$.  As the figure shows, the values in this case form two distinct, compact, non-overlapping clouds, which correspond to the degrees of the nodes~$v$ ``sending'' the messages: the right-most cloud corresponds to nodes of degree~5 while the left-most one corresponds to nodes of degree~10.  This simple observation suggests a possible approximation to the message passing equations, in which we approximate each message with the mean or centroid value of the cloud to which it belongs (shown by the plus symbols in the figure), in effect assuming that the messages are a function of degree only.  This approximation, as we will see, turns out to give remarkably accurate estimates of the spectral density in many cases.

Beyond numerical results like those in Fig.~\ref{fig:messages}, there is some precedent for a degree-based approximation of this kind to be found in the previous literature on complex networks.  In studies of epidemic models, for instance, which can be written in the form of a message passing process~\cite{KN10a}, Pastor-Satorras and Vespignani~\cite{PV01a,PV01b} found a similar degree-based approximation to work well.  The approximation also bears some conceptual similarity to the ``effective medium approximation'' introduced by Semerjian and Cugliandolo~\cite{SC02b} for the Erd\H{o}s--R\'enyi random graph (building on previous work by Biroli and Monasson~\cite{BM99}) and by Dorogovtsev~\etal~\cite{DGMS03} for the configuration model, although the details of the functional forms are different in both cases.

So consider a configuration model with given node degrees such that the fraction of nodes with degree~$k$ is~$p_k$.  Our approximation consists of replacing each message~$h^{uv}(z)$ by the mean value~$h_k(z)$ of messages sent from nodes with the same degree~$k$ as node~$v$.  Rearranging our message passing equation, Eq.~\eqref{eq:messages}, as
\begin{equation}
z^2 h^{uv}(z) = 1 + z^2 h^{uv}(z) \sum_{\substack{w\in N_v\\w\ne u}} h^{vw}(z),
\end{equation}
making the replacement $h^{uv}(z)\to h_k(z)$, and averaging over all edges~$(u,v)$ where $v$ has degree~$k$, of which there are $nkp_k$, we get
\begin{align}
z^2 h_k(z) &= 1 + z^2 h_k(z) {1\over nkp_k} \sum_{v:k_v=k}\,\sum_{u\in N_v}\,
   \sum_{\substack{w\in N_v\\w\ne u}} h^{vw}(z) \nonumber\\
  &= 1 + z^2 h_k(z) {k-1\over nkp_k} \sum_{v:k_v=k}\,\sum_{w\in N_v} h^{vw}(z).
\end{align}
In a configuration model network, however, the degrees of adjacent nodes~$v,w$ are uncorrelated, so the average of $h^{vw}(z)$ over many different~$w$ is independent of~$k$ and, in the limit of large network size, simply equal to the average message in the network as a whole, which we will denote~$h(z)$:
\begin{equation}
{1\over nkp_k} \sum_{v:k_v=k} \sum_{w\in N_v} h^{vw}(z) \to h(z).
\end{equation}
Thus we have $z^2 h_k(z) = 1 + z^2 (k-1) h_k(z) h(z)$ or
\begin{equation}
h_k(z) = {1/z^2\over1 - (k-1) h(z)}.
\end{equation}

Furthermore, the average message~$h(z)$ is itself equal to the average of~$h_k(z)$ over all degrees~$k$, but here we must be careful.  The node~$v$ appearing in the message~$h^{uv}(z)$ is by definition reached by following an edge from node~$u$, and the degrees of nodes reached by following an edge are drawn not from the overall degree distribution~$p_k$ of the network as a whole, but from the modified distribution $kp_k/c$ where $c=\sum_k k p_k$ is the average degree~\cite{NSW01}.  Commonly this is expressed in terms of the so-called \defn{excess degree distribution}:
\begin{equation}
q_k = {(k+1)p_{k+1}\over c},
\end{equation}
which is the probability distribution of the number of edges attached to the node other than the one we followed to reach it.  In terms of this quantity, the average message is given by
\begin{align}
h(z) &= \sum_{k=1}^\infty {kp_k\over c} h_k(z)
      = {1\over cz^2} \sum_{k=1}^\infty {k p_k\over1 - (k-1) h(z)}
        \nonumber\\
     &= {1\over z^2} \sum_{k=0}^\infty {q_k\over1 - k h(z)},
\label{eq:hz}
\end{align}
where we have made the change of variables $k\to k+1$ in the final equality.

We can use this result to calculate the spectral density itself by making a similar degree-based approximation to the quantities~$g^u(z)$ appearing in Eq.~\eqref{eq:gu}.  We assume that $g^u(z)$ is well approximated by the mean~$g_k(z)$ of $g^u(z)$ over all nodes~$u$ with degree~$k$.  Rearranging Eq.~\eqref{eq:gu}, making this approximation, and averaging again over nodes of degree~$k$, we find that $g_k(z) = 1 + k g_k(z) h(z)$, or
\begin{equation}
g_k(z) = {1\over1-kh(z)}.
\end{equation}
Summing over all nodes, we then get
\begin{equation}
\sum_{u=1}^n g^u(z) = \sum_{k=0}^\infty np_k g_k(z)
     = n \sum_{k=0}^\infty {p_k\over1 - kh(z)},
\label{eq:gz}
\end{equation}
and substituting this result into Eq.~\eqref{eq:rho3} we find that, within this approximation, the spectral density is given by
\begin{equation}
\rho(z) = -{1\over\pi z} \sum_{k=0}^\infty {p_k\over1 - k h(z)}.
\label{eq:rho4}
\end{equation}

Between them, Eqs.~\eqref{eq:hz} and~\eqref{eq:rho4} now give us a complete formula for calculating the spectral density.  Notice that, by contrast with the original message passing method, these equations do not depend on the precise form of the network, or even on its size---they are a function only of the degree distribution.  Moreover, the average over messages becomes more and more accurate as network size grows, so that in effect Eqs.~\eqref{eq:hz} and~\eqref{eq:rho4} give us (an approximation to) the spectrum of the configuration model in the limit of large size.

Equation~\eqref{eq:hz} can be solved for $h(z)$ for any given value of~$z$ by simple iteration, starting from a suitable initial value, such as $h(z)=0$.  Then we substitute the result into Eq.~\eqref{eq:rho4} to get~$\rho(z)$.  Note that for each value~$z$ at which we want to calculate the spectral density we need now iterate only one equation, Eq.~\eqref{eq:hz}, in contrast to the full message passing method of Eq.~\eqref{eq:messages}, which involves iterating equations equal in number to twice the number of edges in the network, which can be thousands or millions in real-world situations.  Equation~\eqref{eq:hz} still contains a sum over degrees~$k$, but this sum has only as many terms as there are distinct degrees in the network, which in most cases is far smaller than the number of edges.  Indeed, as we have said the equations given here apply in the limit of infinite size, for which the direct iteration of Eq.~\eqref{eq:messages} would of course be impossible.

\begin{figure}
\begin{center}
\includegraphics[width=\columnwidth]{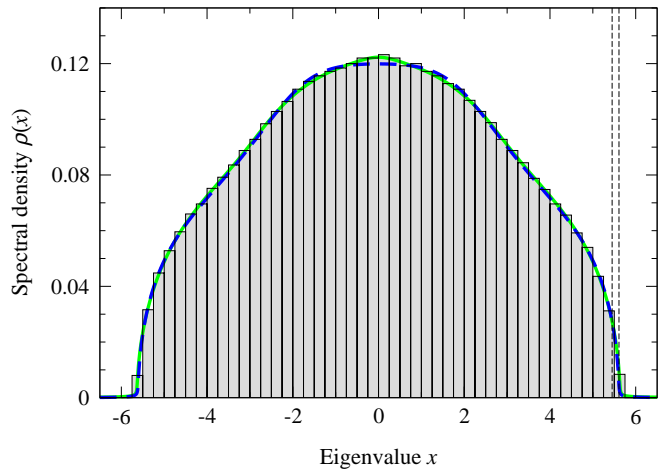}
\end{center}
\caption{The spectral density of a configuration model network in which nodes are randomly assigned one of two different degrees, 5~and~10, with equal probability.  The histogram shows the result of a direct numerical diagonalization of the adjacency matrix of a single such network with $10\,000$ nodes, while the solid curve (green) shows the spectral density for the same single network computed using the message passing approach of Eq.~\eqref{eq:rho3} with $z = x + \ii\epsilon$ and $\epsilon=0.01$.  The dashed curve (blue) shows results from the degree-based approximation proposed in this paper.  The two curves are quite difficult to distinguish because they coincide very closely.  The dashed vertical lines on the right-hand side of the figure denote the upper and lower bounds on the position of the band edge derived in Section~\ref{sec:bandedge}.}
\label{fig:spect-5-10}
\end{figure}

As an example application of our approach consider Fig.~\ref{fig:spect-5-10}, which shows the spectrum of a network drawn from the same model as Fig.~\ref{fig:messages}, in which nodes have just two distinct degrees, 5~and~10, with equal probability.  The histogram in the figure shows the results of a direct numerical diagonalization of the adjacency matrix of a single such network with $10\,000$ nodes, while the solid curve shows the spectral density calculated from the full message passing method, Eq.~\eqref{eq:rho3}.  The dashed curve shows the spectrum calculated using our degree-based approximation, Eqs.~\eqref{eq:hz} and~\eqref{eq:rho4}.  As we can see, the agreement between all three calculations is excellent, and in particular there is barely any perceptible difference between the full and approximate versions of the message passing calculation.  To the extent that the two differ, it is mostly in the center of the figure in the region close to~$x=0$, a pattern that we will see repeated in other examples.

Note moreover that, while the full message passing calculation took several minutes of computer time to complete, the approximate calculation took only a few seconds, essentially all of which was spent on iterating Eq.~\eqref{eq:hz}.  A faster calculation still might be possible by using a more efficient method of solution than simple iteration, such as Newton's method.

As an aside, we note that essentially the same method of calculation can be applied to the Chung--Lu model, the model in which the edges are independent random variables and only the expected degrees of the nodes are fixed, not the actual degrees~\cite{CL02b}.  This model also produces locally tree-like networks, meaning that the message passing method is applicable, and furthermore one can usefully approximate the message passing equations by assuming that messages originating at all nodes with the same expected degree take the same value.  Following a similar argument to that for the configuration model above, we then arrive at the following equations for the spectral density:
\begin{equation}
h(z) = {1\over z^2} \int_0^\infty\! {q_k\>\dd k\over1 - k h(z)},\quad
\rho(z) = -{1\over \pi z} \int_0^\infty\! {p_k\>\dd k\over1 - k h(z)},
\end{equation}
where $k$ now represents not the actual degree of a node but its expected degree, which can take any non-negative real value.

Interestingly, after some translation of notation these are exactly the same equations that were derived previously for the Chung--Lu model using methods from free probability theory~\cite{NN13}.  Superficially, they look identical to Eqs.~\eqref{eq:hz} and~\eqref{eq:rho4} for the configuration model, apart from the replacement of the sums by integrals, but the similarity is deceptive.  The definition of the excess degree distribution~$q_k$ for the Chung--Lu model is different from that for the configuration model.  (For the Chung--Lu model the correct definition is $q_k = k p_k/c$, compared with $q_k = (k+1) p_{k+1}/c$ for the configuration model.)  This can make a substantial difference to the shape of the spectrum and if one simply applies the solution for the Chung--Lu model to the configuration model the results are quite poor, particularly for sparse networks.  Nonetheless, the degree-based approximation does give a more straightforward derivation of the equations for the Chung--Lu model, requiring less advanced techniques than the free probability approach.  (The reverse procedure does not seem to work, however: there is no obvious way to use free probability theory to derive the equations for the configuration model.  The sticking point is that the adjacency matrix and the degree sequence are not ``free'' with respect to one another in the free probability sense.)

\subsection{Analytic solutions}
In some cases Eq.~\eqref{eq:hz} allows a closed-form solution for~$h(z)$, depending on the form of the degree distribution.  As a simple example consider a regular network, meaning one in which all nodes have the same degree~$c$, for which Eq.~\eqref{eq:hz} takes the form
\begin{equation}
h(z) = {1/z^2\over1 - (c-1) h(z)}.
\end{equation}
This can be rearranged into the form of a quadratic equation $(c-1)h^2 - h + 1/z^2 = 0$, with solutions
\begin{equation}
h(z) = {1 \pm \sqrt{1-4(c-1)/z^2}\over 2(c-1)}.
\end{equation}
Then
\begin{equation}
\rho(z) = {1\over \pi z[ch(z)-1]},
\end{equation}
and, after taking the imaginary part and going to the real line, we find the spectral density to be
\begin{equation}
\rho(x) = (c/2\pi) {\sqrt{4(c-1)-x^2}\over c^2-x^2},
\end{equation}
which recovers the standard Kesten--McKay distribution for a random regular graph~\cite{McKay81}.  Thus in this case our approximation is not an approximation at all: the Kesten--McKay distribution gives the exact spectral density of a random regular graph in the limit of large size.  The calculation is exact because in a regular graph the neighborhood of every node has the same network structure, so all messages are in fact exactly equal.

For a slightly more complex example consider again a network of the kind in Fig.~\ref{fig:spect-5-10}, with nodes of two different degrees, which we will denote~$a$ and~$b$.  For such a network Eq.~\eqref{eq:hz} becomes
\begin{equation}
h(z) = {1/z^2\over ap_a+bp_b} \biggl[ {ap_a\over1-(a-1)h(z)}
       + {bp_b\over1-(b-1)h(z)} \biggr],
\end{equation}
which gives a cubic equation for~$h(z)$, which is solvable in closed form though the solution is complicated and we will not reproduce it here.  Applied, for instance, to the example network in Fig.~\ref{fig:spect-5-10}, it gives essentially the same result as our numerical solution for the same system.

More broadly, we can define a moment generating function~$\mu_p(x)$ for a distribution~$p_k$ to be the power series in~$x$ whose coefficients are the moments~$\av{k^r}_p$ of~$p_k$ thus:
\begin{equation}
\mu_p(x) = \sum_{r=0}^\infty \av{k^r}_p x^r
   = \biggl\langle {1\over1-kx} \biggr\rangle_{\!\!p}
   = \sum_{k=0}^\infty {p_k\over1-kx}.
\end{equation}
In terms of such generating functions, the fundamental equation~\eqref{eq:hz} can be written
\begin{equation}
h(z) = {1\over z^2}\,\mu_q(h(z)),
\end{equation}
and Eq.~\eqref{eq:rho4} becomes
\begin{equation}
\rho(z) = -{1\over\pi z}\,\mu_p(h(z)).
\end{equation}
If one knows the moment generating functions for a particular degree distribution, one can use these equations to derive the spectral density.

\subsection{Position of the band edges}
\label{sec:bandedge}
One issue of particular interest is the position of the edges of the band of nonzero spectral density in the spectrum of a network.  The upper band edge plays a role for instance in locating localization transitions in networks~\cite{MZN14} and the position of the so-called detectability threshold for community detection~\cite{DKMZ11a,NN12}.

\begin{figure}
\begin{center}
\includegraphics[width=\columnwidth]{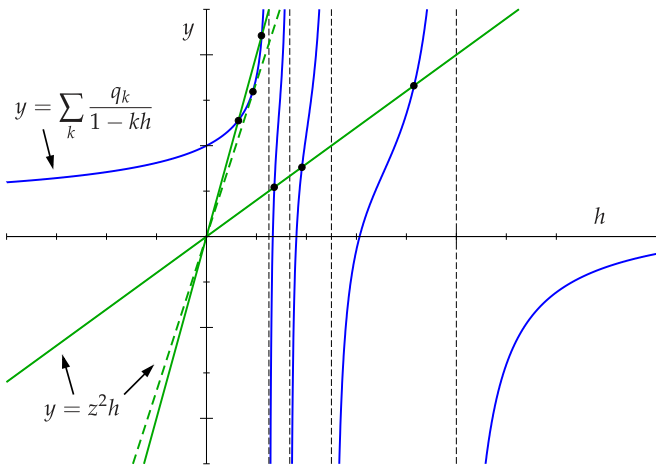}
\end{center}
\caption{Graphical representation of the solution of Eq.~\eqref{eq:solution}.  The curves represent the right side of the equation, while the three diagonal lines represent three possible values of the left side.  Where curve and line cross, marked by the dots, are the real solutions of the equation.  If the diagonal line intersects the left-most curve segment then all solutions are real.  If it does not, however, there will be two complex solutions, which give rise to a nonzero spectral density.}
\label{fig:solution}
\end{figure}

To understand why there is a finite band at all, and where its edges fall, consider Fig.~\ref{fig:solution}, which sketches a graphical solution to the fundamental equation~\eqref{eq:hz} for~$h(z)$.  We rearrange the equation in the form
\begin{equation}
z^2 h = \sum_{k=0}^\infty {q_k\over1-kh},
\label{eq:solution}
\end{equation}
and then separately plot the left and right sides of the equation as a function of~$h$.  Where the two cross are our solutions for~$h(z)$.

The right-hand side of the equation, represented by the rising curves in Fig.~\ref{fig:solution}, has simple poles at $h = 1/k$ for all nonzero values~$k$ of the excess degree that occur in the network.  In the figure we assume that values $k=0\ldots4$ occur, so that there are poles at $h=1$, $\frac12$, $\frac13$, and~$\frac14$.  Meanwhile, the left-hand side~$z^2h$ takes the form of a simple straight line through the origin with slope~$z^2$.  Lines are sketched in the figure for three different values of~$z^2$.  The solutions of~\eqref{eq:solution} are marked with dots.

Generically, Eq.~\eqref{eq:solution} gives us a polynomial equation of degree~$m+1$, where $m$ is the number of distinct nonzero values of the excess degree, or equivalently the number of poles in the right-hand side of the equation.  This in turn means the equation has $m+1$ solutions for~$h(z)$.  As we can see from the geometry of Fig.~\ref{fig:solution} there are always at least $m-1$ real solutions, one in each of the intervals between the $m$ poles.  Real solutions, however, cannot give us a nonzero spectral density---we need complex $h(z)$ to get a nonzero density when we take the imaginary part of Eq.~\eqref{eq:rho4}.  Thus we focus on the remaining two solutions, which can be either real or complex, with the band edge corresponding to the point at which complex solutions first emerge.

There are three different possible forms of the solution, corresponding to the three different values of~$z^2$ depicted in Fig.~\ref{fig:solution}.  If $z^2$ is sufficiently large (the steepest line in the figure), the line~$z^2h$ intersects the first segment of the curve representing the right-hand side of Eq.~\eqref{eq:solution}, giving us two real solutions for~$h(z)$.  If this happens then all $m+1$ solutions for~$h$ are real, there are no complex solutions at all, and the spectral density is zero.  In this regime we are outside the spectral band.  Conversely, if $z^2$ is sufficiently small, then the line of $z^2h$ does not intersect the first segment, there are only $m-1$ real solutions, and the two remaining solutions must necessarily be complex, placing us inside the band.  Between these two regimes lies the borderline case---the band edge---represented by the dashed diagonal line in the figure, which is precisely tangent to the first segment of the curve.

While it is difficult, or sometimes impossible, to solve exactly for the roots of polynomial equations, we can derive useful bounds on the position of the band edge by inspecting the geometry of Fig.~\ref{fig:solution}.  Note that the curve representing the right-hand side of Eq.~\eqref{eq:solution} intercepts the vertical axis at $y=\sum_k q_k = 1$, since $q_k$ is a properly normalized probability distribution.  Thus the tangent point must fall at $y>1$.  At the same time the horizontal coordinate of the tangent point must satisfy $h<1/K$, where $K$ is the largest value of the excess degree~$k$ in the network (which is 1 less than the largest value of the ordinary degree).  Thus the critical slope of the tangent line at the band edge satisfies
\begin{equation}
z^2 = {y\over h} > {1\over1/K} = K,
\end{equation}
meaning that the upper band edge falls at a point $z \ge \sqrt{K}$ and the lower one falls at $z\le-\sqrt{K}$.  For the network in Fig.~\ref{fig:spect-5-10}, for example, which has nodes of degree 5 and 10 only, the largest excess degree is $K=9$, and hence the upper band edge satisfies $z \ge 3$.  From an inspection of Fig.~\ref{fig:spect-5-10} this appears to be correct---the band edge looks to fall at around $z=5.5$.  It is not a very good bound, though it is interesting nonetheless, since it implies that in a network in which node degrees are unbounded there will be no upper edge to the eigenvalue spectrum: the band edge diverges as $K$ diverges.  In a network with an exponential or power-law degree distribution, for instance, there will be no upper limit to the spectral band.

We can derive a better bound on the position of the band edge by first computing a lower bound on the right-hand side of~\eqref{eq:solution} in the region $h<1/K$ thus:
\begin{align*}
\sum_{k=0}^\infty {q_k\over1-kh} &=
     {q_K \over 1-Kh} + \sum_{k(\ne K)} {q_k\over1-kh} \nonumber\\
  &\ge {q_K \over 1-Kh} + \sum_{k(\ne K)} q_k \nonumber\\
  &= {q_K \over 1-Kh} + 1 - q_K,
\label{eq:bound}
\end{align*}
and the value of $z^2$ at the band edge is always greater than the slope of the tangent line to this curve.  The tangent falls at the point where there is a double root of
\begin{equation}
z^2 h = {q_K \over 1-Kh} + 1 - q_K,
\label{eq:boundapprox}
\end{equation}
which is equivalent to the quadratic equation
\begin{equation}
Kz^2h^2 - [(1-q_K)K+z^2] h + 1 = 0.
\label{eq:quadratic}
\end{equation}
This has a double root when its discriminant vanishes, which gives us another quadratic equation $z^4 - 2(1+q_K)Kz^2 + (1-q_K)^2K^2 = 0$ for~$z^2$, whose solution now gives us an improved bound on the position of the band edge:
\begin{equation}
z \ge \sqrt{K}\bigl(1+\sqrt{q_K}\bigr).
\end{equation}
Taking the example of our network with equal fractions of nodes of degree~5 and 10 again, so that $K=9$ and $q_K=\frac23$, we find that $z\ge 3(1+\sqrt{2/3}) \simeq 5.449$.

\begin{figure*}
\begin{center}
\includegraphics[width=13.2cm]{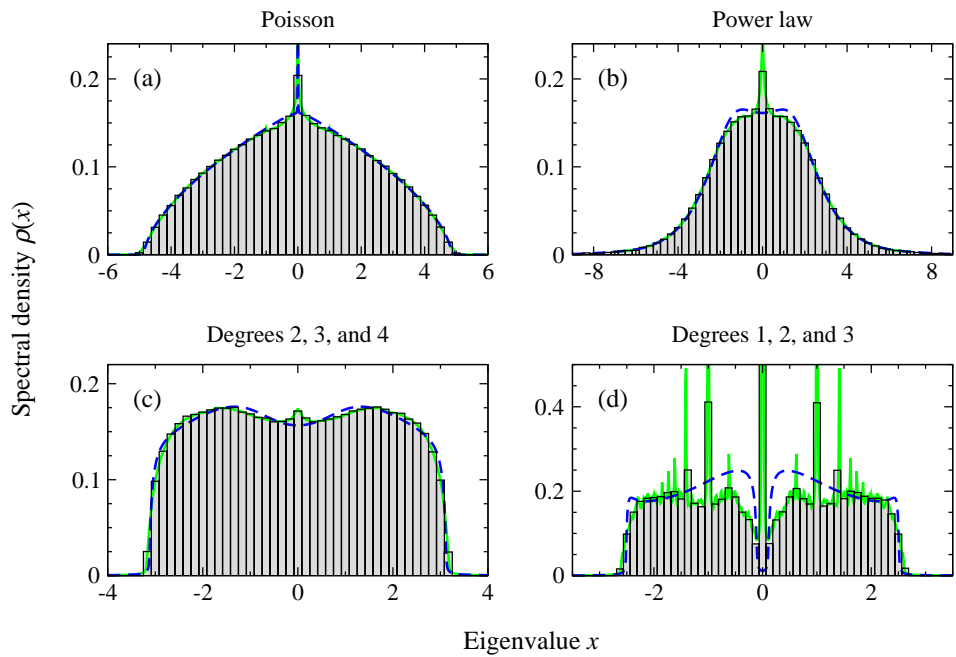}
\end{center}
\caption{Each panel shows the spectrum of a configuration model with one particular choice of degree distribution.  In each case the histogram shows the spectrum of a single realization of the model with $10\,000$ nodes, the solid (blue) curve shows the spectrum of the same single realization computed using the message passing method of Section~\ref{sec:messages} with $\epsilon=0.01$, while the dashed (green) curve shows the degree-based approximation of Section~\ref{sec:approximation}, also with $\epsilon=0.01$.  The four degree distributions are: (a)~Poisson degree distribution with mean~5; (b)~power-law degree distribution generated using the model of Barab\'asi and Albert~\cite{BA99b} with parameter~$m=3$; (c)~nodes of degree 2, 3, and~4, with equal probability; (d)~nodes of degree 1, 2, and~3 with equal probability.  For (a) and (b) we use the degree distribution of the actual network in our degree-based approximation, not the formal distribution from which the degrees were drawn.}
\label{fig:examples}
\end{figure*}

We can also derive an upper bound on the position of the band edge by rearranging Eq.~\eqref{eq:solution} into the form
\begin{equation}
z^2 = {1\over h} \sum_{k=0}^\infty {q_k\over1-kh}.
\end{equation}
For any real $h<1/K$ this gives an upper bound on the value of~$z^2$ at the band edge, with values of $h$ closer to the true double root giving better bounds.  We use the value of $h$ at the double root of the approximation, Eq.~\eqref{eq:boundapprox}, which is $h = 1/[K(1+\sqrt{q_K})]$.  For our network with nodes of degree 5 and~10 this yields an upper bound of $5.609$ on the position of band edge, meaning that the band edge falls in the relatively narrow interval $5.449\le z \le 5.609$.  This interval is shown on Fig.~\ref{fig:spect-5-10} by the dashed vertical lines, and appears to agree well with both the exact position of the band edge from the full message passing calculation and the approximate position derived from the degree-based approximation.

\section{Examples}
\label{sec:examples}
In this section we look at some applications of our methods to example networks, illustrating the advantages and limitations of the approach.

Figure~\ref{fig:examples} shows spectra for four networks generated from configuration models with different degree distributions.  Panel~(a) shows the spectrum of a network with a Poisson degree distribution with mean degree~$c=5$---effectively a standard Erd\H{o}s--R\'enyi style random graph in which each possible edge is present with the same probability~$c/n$.  Since the elements of the adjacency matrix in such a network are independent identically-distributed random variables, one might expect the spectrum to follow the standard Wigner semicircle distribution.  However, as shown previously by many authors~\cite{EJ76,FDBV01,SC02b,Kuhn08}, this is not the case when, as here, the graph is very sparse, meaning $c$ is small.  The deviation from the Wigner law is clear in Fig.~\ref{fig:examples}a, with the spectrum having a distinctly non-semicircular shape with a peak at $x=0$.  The histogram in the figure shows the distribution of eigenvalues calculated by direct diagonalization of a single instance of the model with $n=10\,000$ nodes, while the solid curve shows the spectrum calculated using the full message passing method of Section~\ref{sec:messages}.  The dashed curve shows the results of the degree-based approximation introduced here and, as we can see, it works well in this case, being barely distinguishable from the full calculation.

The Poisson distribution, however, is not a good approximation to the degree distributions of most real-world networks, which are typically strongly right-skewed~\cite{BA99b,ASBS00}.  Many networks are observed to have degree distributions that approximately follow a power law or Pareto distribution.  Figure~\ref{fig:examples}b shows the spectrum for such a network, which displays characteristic long tails with no clear band edge, as noted previously for instance in~\cite{GKK01a}, where it is shown that the leading eigenvalue of a power-law network scales as $n^{1/4}$ with system size, and is as a result unbounded in the limit of large~$n$.  Again the degree-based approach does a good job of approximating the spectrum of the network, although there are now clear deviations visible for small absolute values of~$x$ (i.e.,~close to the origin).  In particular, notice that, by contrast with panel~(a), the peak at $x=0$ is not well reproduced by our approximation (but is captured by the full message passing calculation).

In general, while we find that the degree-based approximation does well in many cases, it is weakest when the network is particularly sparse.  In panel~(c) of the figure, for example, we show results for a network with nodes of degree 2, 3, and~4 only, in equal numbers.  This network is now very sparse---the average degree is only~3---and differences between the true spectrum (solid line) and approximation (dashed line) are becoming visible.

Panel~(d) shows a particularly extreme case, of a network with nodes of degree 1, 2, and~3 only for which the spectrum becomes quite ill-behaved with notable spikes and other irregularities.  While these are once again reproduced faithfully by the full message passing calculation, our approximation fails to capture them, and moreover does a relatively poor job of the overall shape of the spectrum.  The spikes in the spectrum are due to the presence of numerous small components in the network.  The peaks at $\pm1$ are due to components of size~2, for example.  Our approximation fails to pick these out because it does not distinguish between messages in small components and those in the giant component.  The method does, however, still capture the basic shape of the spectrum further from the origin, and gets the positions of the band edges approximately correctly.

\section{Conclusions}
\label{sec:concs}
In this paper we have derived a set of message passing equations that allow one to calculate the spectrum of the adjacency matrix of an arbitrary, tree-like network.  For the particular case of the configuration model, we approximate these equations by assuming that the messages are a function of node degree only, which gives a much simplified form for the spectral density that applies in the limit of large network size.  Test applications of these methods on a range of networks shows that the approximation works well for all but the sparsest of networks, and where it does show deviations from the true spectral density it is typically in the region close to the origin and away from the band edges.  We have also used our approximate equations as a starting point for deriving bounds on the positions of the edges of the spectrum, and in particular the upper edge of the continuous spectral band, which plays a role in determining the locations of structural phase transitions in networks.

It is interesting to ask if there are better approximations we could make to the message passing equations than the one introduced here, and there are a number of possibilities.  One could imagine an approximation where the clouds of points appearing in Fig.~\ref{fig:messages} are represented by more than one different value: a large cloud could be divided into two parts, each of which is approximated by its own individual centroid.  Another possibility is that we could attempt to represent the messages for the lowest degree nodes more accurately in some way.  Most of the error in our approximation is in the low-degree nodes, since these are the ones for which the sum of incoming messages will show the largest statistical fluctuation.  It is for this reason that the approximation works poorly for very sparse networks, such as the one in Fig.~\ref{fig:examples}d.  One could imagine, for instance, explicitly summing over the possible degrees of the neighbors of a low-degree node, producing a kind of two-step approximation for these nodes that would presumably be more accurate than the one-step approach we currently employ.  These possibilities, however, we leave for future work.

\bigskip\begin{acknowledgments}
This work was funded in part by the NSF under grants DMS--1407207 and DMS--1710848 (MEJN) and by ONR grant N00014-15-1-2141, DARPA Young Faculty Award D14AP00086, and ARO MURI grants W911NF-11-1-0391 and 2015-05174-05 (RRN).
\end{acknowledgments}


\begin{thebibliography}{10}
\expandafter\ifx\csname url\endcsname\relax
  \def\url#1{\texttt{#1}}\fi
\expandafter\ifx\csname urlprefix\endcsname\relax\def\urlprefix{URL }\fi

\bibitem{Bonacich87}
P.~F. Bonacich, Power and centrality: A family of measures. \textit{Am. J.
  Sociol.} \textbf{92}, 1170--1182 (1987).

\bibitem{Fiedler73}
M.~Fiedler, Algebraic connectivity of graphs. \textit{Czech. Math. J.}
  \textbf{23}, 298--305 (1973).

\bibitem{PSL90}
A.~Pothen, H.~Simon, and K.-P. Liou, Partitioning sparse matrices with
  eigenvectors of graphs. \textit{SIAM J. Matrix Anal. Appl.} \textbf{11},
  430--452 (1990).

\bibitem{Newman06b}
M.~E.~J. Newman, Modularity and community structure in networks. \textit{Proc.
  Natl. Acad. Sci. USA} \textbf{103}, 8577--8582 (2006).

\bibitem{PG16}
M.~A. Porter and J.~Gleeson, \textit{Dynamical Systems on Networks: A
  Tutorial}. Springer, Berlin (2016).

\bibitem{BBCR10}
B.~Bollob\'as, C.~Borgs, J.~Chayes, and O.~Riordan, Percolation on dense graph
  sequences. \textit{Annals of Probability} \textbf{38}, 150--183 (2010).

\bibitem{KNZ14}
B.~Karrer, M.~E.~J. Newman, and L.~Zdeborov\'a, Percolation on sparse networks.
  \textit{Phys. Rev. Lett.} \textbf{113}, 208702 (2014).

\bibitem{CM11}
M.~Cucuringu and M.~W. Mahoney, Localization on low-order eigenvectors of data
  matrices. Preprint arxiv:1109.1355 (2011).

\bibitem{MZN14}
T.~Martin, X.~Zhang, and M.~E.~J. Newman, Localization and centrality in
  networks. \textit{Phys. Rev. E} \textbf{90}, 052808 (2014).

\bibitem{DKMZ11a}
A.~Decelle, F.~Krzakala, C.~Moore, and L.~Zdeborov\'a, Inference and phase
  transitions in the detection of modules in sparse networks. \textit{Phys.
  Rev. Lett.} \textbf{107}, 065701 (2011).

\bibitem{NN12}
R.~R. Nadakuditi and M.~E.~J. Newman, Graph spectra and the detectability of
  community structure in networks. \textit{Phys. Rev. Lett.} \textbf{108},
  188701 (2012).

\bibitem{FDBV01}
I.~J. Farkas, I.~Der\'enyi, A.-L. Barab\'asi, and T.~Vicsek, Spectra of
  ``real-world'' graphs: Beyond the semicircle law. \textit{Phys. Rev. E}
  \textbf{64}, 026704 (2001).

\bibitem{GKK01a}
K.-I. Goh, B.~Kahng, and D.~Kim, Spectra and eigenvectors of scale-free
  networks. \textit{Phys. Rev. E} \textbf{64}, 051903 (2001).

\bibitem{DGMS03}
S.~N. Dorogovtsev, A.~V. Goltsev, J.~F.~F. Mendes, and A.~N. Samukhin, Spectra
  of complex networks. \textit{Phys. Rev. E} \textbf{68}, 046109 (2003).

\bibitem{SC02b}
G.~Semerjian and L.~F. Cugliandolo, Sparse random matrices: The eigenvalue
  spectrum revisited. \textit{J. Phys. A} \textbf{35}, 4837--4852 (2002).

\bibitem{EJ76}
S.~F. Edwards and R.~C. Jones, The eigenvalue spectrum of a large symmetric
  random matrix. \textit{J. Phys. A} \textbf{9}, 1595--1603 (1976).

\bibitem{Kuhn08}
R.~K{\"u}hn, Spectra of sparse random matrices. \textit{J. Phys. A}
  \textbf{41}, 295002 (2008).

\bibitem{RCKT08}
T.~Rogers, I.~P\'erez~Castillo, R.~K{\"u}hn, and K.~Takeda, Cavity approach to
  the spectral density of sparse symmetric random matrices. \textit{Phys. Rev.
  E} \textbf{78}, 031116 (2008).

\bibitem{CLV03}
F.~Chung, L.~Lu, and V.~Vu, Spectra of random graphs with given expected
  degrees. \textit{Proc. Natl. Acad. Sci. USA} \textbf{100}, 6313--6318 (2003).

\bibitem{CL02b}
F.~Chung and L.~Lu, The average distances in random graphs with given expected
  degrees. \textit{Proc. Natl. Acad. Sci. USA} \textbf{99}, 15879--15882
  (2002).

\bibitem{NN13}
R.~R. Nadakuditi and M.~E.~J. Newman, Spectra of random graphs with arbitrary
  expected degrees. \textit{Phys. Rev. E} \textbf{87}, 012803 (2013).

\bibitem{MR95}
M.~Molloy and B.~Reed, A critical point for random graphs with a given degree
  sequence. \textit{Random Structures and Algorithms} \textbf{6}, 161--179
  (1995).

\bibitem{NSW01}
M.~E.~J. Newman, S.~H. Strogatz, and D.~J. Watts, Random graphs with arbitrary
  degree distributions and their applications. \textit{Phys. Rev. E}
  \textbf{64}, 026118 (2001).

\bibitem{KN10a}
B.~Karrer and M.~E.~J. Newman, A message passing approach for general epidemic
  models. \textit{Phys. Rev. E} \textbf{82}, 016101 (2010).

\bibitem{PV01a}
R.~Pastor-Satorras and A.~Vespignani, Epidemic spreading in scale-free
  networks. \textit{Phys. Rev. Lett.} \textbf{86}, 3200--3203 (2001).

\bibitem{PV01b}
R.~Pastor-Satorras and A.~Vespignani, Epidemic dynamics and endemic states in
  complex networks. \textit{Phys. Rev. E} \textbf{63}, 066117 (2001).

\bibitem{BM99}
G.~Biroli and R.~Monasson, A single defect approximation for localized states
  on random lattices. \textit{J. Phys. A} \textbf{32}, L255--L261 (1999).

\bibitem{McKay81}
B.~D. McKay, The expected eigenvalue distribution of a large regular graph.
  \textit{Linear Algebra Appl.} \textbf{40}, 203--216 (1981).

\bibitem{BA99b}
A.-L. Barab\'asi and R.~Albert, Emergence of scaling in random networks.
  \textit{Science} \textbf{286}, 509--512 (1999).

\bibitem{ASBS00}
L.~A.~N. Amaral, A.~Scala, M.~Barth\'elemy, and H.~E. Stanley, Classes of
  small-world networks. \textit{Proc. Natl. Acad. Sci. USA} \textbf{97},
  11149--11152 (2000).

\end{thebibliography}
\end{document}